# Inverse design of dielectric metasurface by spatial coupled mode theory

Zhicheng Wu[1], Xiaoyan Huang[2], Nanfang Yu[2], Zongfu Yu[1]

[1]Department of Electrical and Computer Engineering, University of Wisconsin-Madison, Madison, WI 53706, USA
[2]Department of Applied Physics and Applied Mathematics, Columbia University, New York, NY 10027, USA.

**Abstract**:

Modeling metasurfaces with high accuracy and efficiency is challenging because they have features smaller than the wavelength but sizes much larger than the wavelength. Full wave simulation is accurate but very slow. Popular design paradigms like locally periodic approximation (LPA) reduce the computational cost by neglecting, partially or fully, near-field interactions between meta-units and treating them in an isolated manner. The coupling between meta units has been fully considered by applying temporal coupled mode theory to model the metasurface. However, this method only works for resonance-based metasurfaces. To model the broadly studied dielectric metasurfaces based on the propagation of guided modes, we propose to model the whole system using spatial coupled mode theory (SCMT) where the dielectric metasurface can be viewed as an array of truncated waveguides. An inverse design routine based on this model is then devised and applied to gain improvements over LPA in several scenarios, such as high numerical aperture lens, multi-wavelength focusing, and suppression of coma aberrations. With its accuracy and efficiency, the proposed framework can be a powerful tool to improve the performance of dielectric metasurface on various tasks.

1. **Introduction**

Metasurfaces, or a subwavelength array of scatterers with engineered optical responses, have gained much attention in recent years. By carefully designing the geometries of individual elements (or, meta-units), various optical functionalities have been demonstrated for metasurfaces such as ultrathin lenses, holography, beam steering, and demultiplexing [1]–[6]. Their unique advantages such as the ability to shape the wavefront by design with subwavelength resolution, a micrometer scale form factor and CMOS-compatible fabrication make them promising alternatives to traditional bulky optical elements.

Metasurfaces are typically made of a large array of sub-wavelength scatterers, whose shape and sizes need to be optimized to achieve specific functionality. Towards this end, various approaches have been proposed with a desire for both high accuracy and low computational costs. To achieve high performance by accurately predicting optical responses, the inverse design of the metasurface based on full-wave simulations has been studied. These routines consist of a forward problem, where a Maxwell solver is used to predict the scattering by the meta-units, and a backward problem, where numerical optimization techniques are used to optimize the meta-surface design based on its performance. The full-wave simulations make no assumptions about the geometry and layout of meta-units and are capable of accurately predicting the phase and amplitude response of arbitrary meta-surfaces, but they require a huge computational load that quickly becomes forbidding as the scale of the meta-surface goes up. There are many attempts to reduce this cost. The commonly used "locally periodic" approximation (LPA) [7] assumes the local variations of meta-unit geometries and hence their optical responses are small so that the scattered field can be approximated by an assembly of scattered fields by individual meta-units of the surface. In this case, their responses are taken to be the same as those of infinite periodic arrays of their replicas, which can be conveniently simulated using finite difference time domain (FDTD) or rigorous coupled wave analysis (RCWA) algorithms, and the entire layout can be designed with ease by picking meta-units with the least local phase errors. This method, while extremely fast, loses accuracy in scenarios where large phase gradients are present (like high NA lenses). One major reason for the decreased accuracy of LPA is that it fails to fully account for the evanescent wave coupling between a meta-unit and its neighbors. Recently, an inverse design paradigm has been proposed based on the temporal coupled mode theory [8]. In this work, the authors model the meta-units as tiny resonators, and their coupling can be treated analytically, powering a fast but accurate numerical design approach. However, this method is limited to meta-surfaces based on resonances and is unfit to model the widely studied dielectric meta-surfaces based on the propagation of guided modes [9].

In this paper, we propose an inverse design paradigm based on the spatial coupled mode theory (SCMT) [10], [11]. The meta-units are modeled as truncated waveguides and their supported modes are calculated. Contrary to the LPA approach, the coupling coefficients between neighboring meta-units are analytically calculated, and together they correctly account for the near-field coupling effects in a meta-surface. On the other hand, this approach is very computationally efficient, being several orders of magnitudes faster than full-wave simulations. We implemented an inverse design paradigm and demonstrated significant improvements over LPA in several scenarios, including a high NA lens and a multi-wavelength lens in the visible. Furthermore, to show that our model works for arbitrary input fields, we minimize the coma aberration of a singlet lens by considering light incident at angles up to 40 degrees. Our results are benchmarked and validated by full-wave simulations. We believe this approach holds great potential in designing large-scale, high-performance meta-surfaces.

2. **Forward solver**

2.1 *SCMT model of metasurface*

We will briefly describe how to calculate the far field of the metasurface given an arbitrary input field, using coupled mode theory. The detailed mathematical models are described in Appendix 9.1 – 9.3. As shown in figure 1, the input field is coupled into the meta-units seen as a truncated waveguide, where only the guided modes are considered. Subsequently, light propagating through the metasurface is approximated as a superposition of all the supported waveguide modes, and they are allowed to interact with modes in neighboring waveguides according to their coupling coefficients. At the exit of the waveguides, the modes are coupled back to free space to obtain the near field profile of the metasurface (the field just above the metasurface). Finally, the near field is propagated freely to the focal plane given a propagation distance, using the Rayleigh-Sommerfeld diffraction theory [12]. Without losing generality, we assume the incident light is S-polarized (perpendicular to the plane of incidence). As a result, only TE modes in the waveguide will be excited. Adding P polarization only doubles the computational cost.

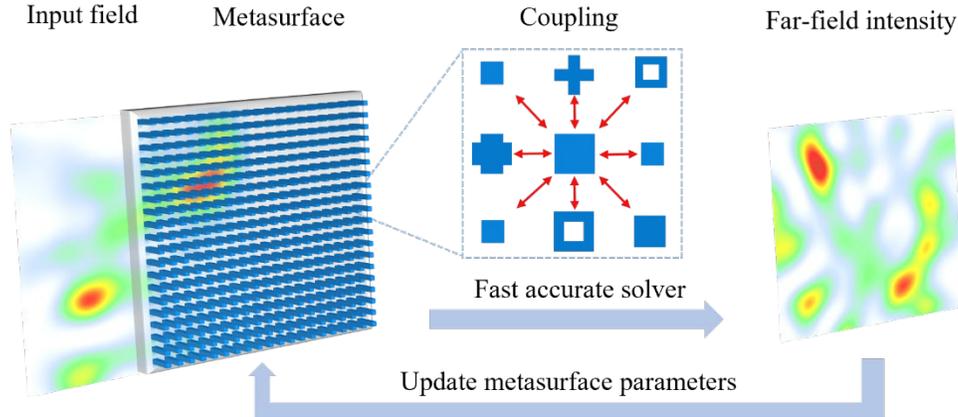

Figure 1. The diagram of the inverse design metasurface by SCMT. The metasurface is treated as an array of waveguides. The process of calculating the scattering field of a metasurface given an arbitrary input field is: First the input field is coupled into the waveguides. Then the field propagation within waveguides is modeled using SCMT which fully captures the coupling between meta units. At the exit of the waveguides, the modes field are coupled back to free space to get the near field. Finally, the near field is transformed into the far field using the Rayleigh-Sommerfeld diffraction equation. Based on the forward simulation, the hyperparameters of the metasurface can be updated depending on the task.

2.2 *Full-wave validation*

We use FDTD simulation to benchmark the accuracy of the SCMT model. First, we use FDTD and SCMT to simulate the near field of a metalens ($\lambda = 650nm$, NA = 0.8, period = 240nm). The comparison of SCMT prediction and the FDTD simulation is shown in figure 2 (a) and (b). Figure (a) shows the comparison of the near field phase. We don't show the near field amplitude because it is nearly constant for the metalens. (b) shows the comparison of far-field intensity. We can see that the SCMT results match very well with the full-wave simulation. When using the LPA method to design a metasurface, people have to assume a plane wave with a fixed incident angle. This is because they do FDTD simulations for all possible meta units with periodic boundary conditions and all simulations should have the same plane wave input. Unlike the LPA method, our model can simulate the response of the metasurface under an arbitrary input field. Because oblique incidence is particularly interesting, as an example, figure (c) (d) shows the near field phase and the focal plane intensity comparison of SCMT prediction and the FDTD simulation with a 15° incident plane wave. The full accuracy analysis of our method compared with the LPA method is provided in Appendix 9.4.

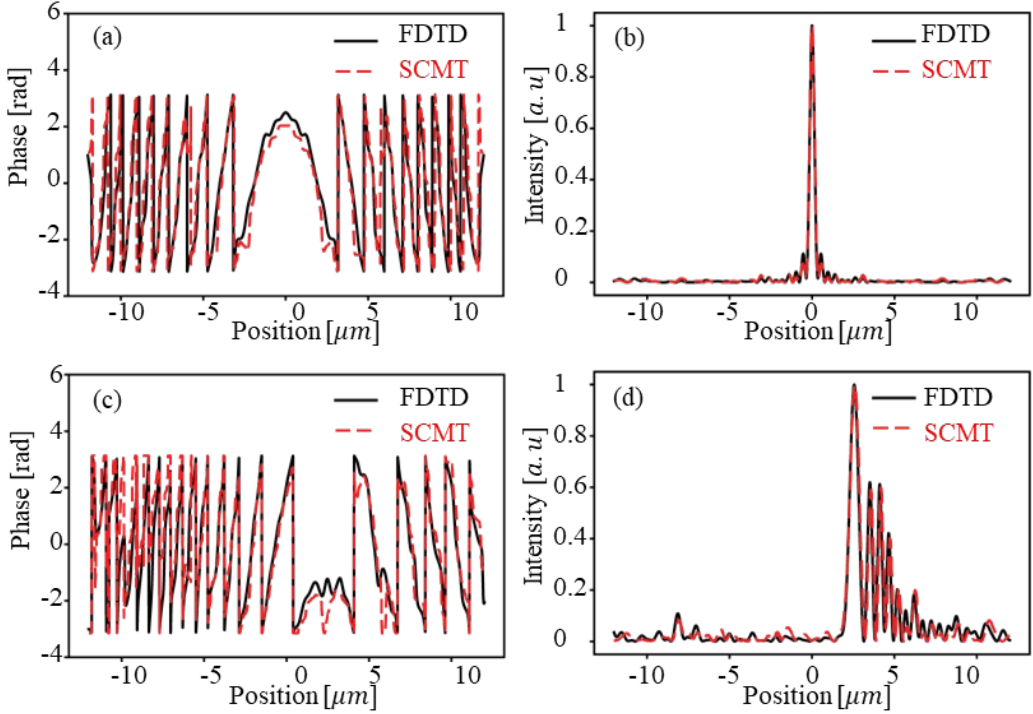

Figure 2. The accuracy analysis of the SCMT method was validated by FDTD. With a normal incident plane wave, (a), and (b) show the near field phase and the focal plane intensity comparison of SCMT prediction and the FDTD simulation. The intensity is normalized by the focal spot intensity. (c), (d) show the phase and intensity comparison of SCMT prediction and the FDTD simulation with a 15° incident plane wave.

## 3. Optimization

In the forward solver, ultimately the far-field intensity can be viewed as a nonlinear function of waveguide widths $w$ given the input field as shown in figure 3. We write the output intensity as:

$$I_{far} = f(w_1, w_2, \ldots w_n | E_{in}) \qquad (1)$$

We want the far-field intensity $I_{far}$ to be as close to a target intensity profile $I_{target}$ as possible. As such, a loss function (object function) can be defined as the L2 distance between them:

$$L = \left\lVert I_{far} - I_{target} \right\rVert^2 \qquad (2)$$

As a result, a metasurface design problem can be viewed as an optimization problem:

$$w_{opt} = \underset{w}{\mathrm{argmin}} \left\lVert f(E_{in}) - I_{target} \right\rVert^2 \qquad (3)$$

Where $w = [w_1, w_2, \ldots w_n]$.

There are many off-the-shelf gradient-based optimization methods to solve this problem [13]–[15]. The key here is to accurately and efficiently calculate the gradient $\frac{\partial L}{\partial w}$. However, the dependence of $I_{far}$ on $w$ is very complex and can't be expressed explicitly in an analytical form. To overcome this problem, we use fully connected neural networks to fit the sub-functions within the model, as shown in figure 3. The details of the fitting are provided in Appendix 9.7. Since our model is fully differentiable, instead of using the adjoint method, we use Pytorch [16] to automatically calculate the gradient after building the forward model, and we employ the Adam optimization algorithm [14] to optimize the meta-surface layout.

Besides supporting gradient-based optimization, the memory usage of the solver is also optimized. Considering a metasurface with $N \times N$ meta units, the size of the coupling matrices $C$ and $K$ is $N^2 \times N^2$ which in theory consume

$O(N^4)$ memory. Because the coupling between two meta units with a distance larger than 2 periods is negligible, we can safely set $C_{ij}, K_{ij} = 0$, where $|i - j| > 2$. As a result, the matrix $B, C, K$ can be represented by sparse matrices in implementation which only consume $O(N^2)$ memory. Although the inverse of a sparse matrix is in general dense, the inverses of $C$ matrix have a very good property that the value of elements will be close to zero when they are away from the diagonal. We managed to represent $C^{-1}$ with $O(N^3)$ memory without compromising accuracy. The details are provided in Appendix 9.8.

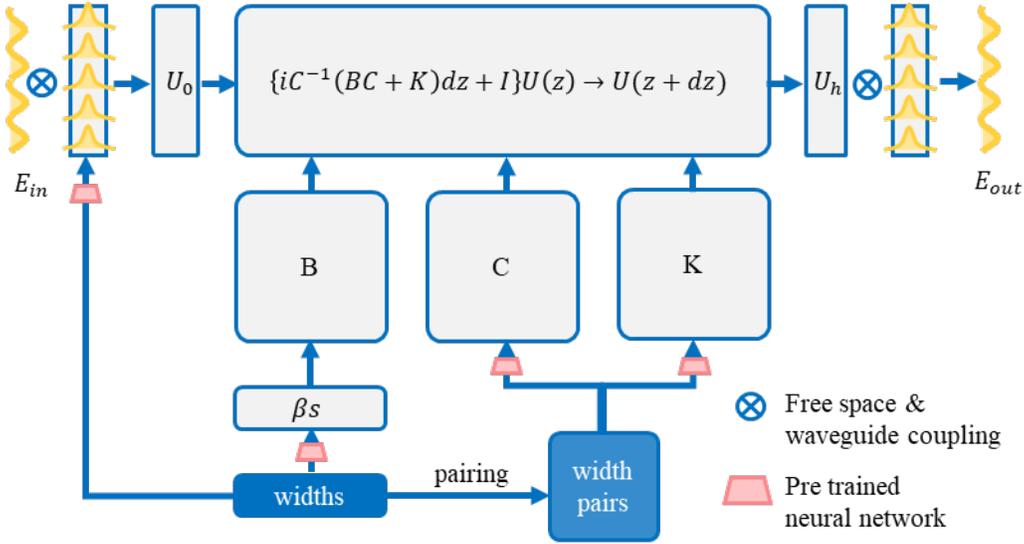

Figure 3. The hybrid architecture of the SCMT model. The forward solver can be viewed as a nonlinear function with the output: near field $E_{out}$ and the input: waveguide widths $w$ and the incident field $E_{in}$. $\beta s$ is an array of propagation constants for each waveguide. The $B$ is a diagonal matrix with $\beta$ as the diagonal element. $C$ and $K$ are coupling matrices. The definition of each item is described in appendix 9.1. The calculation of $\beta_i, c_{ij}, k_{ij}$ replaced by a pre-trained neural network to improve efficiency and support calculating the gradient of widths. The modes propagation within waveguides is differential calculated using the Euler method instead of using the analytic solution provided in appendix 9.1 to save memory.

## 4. Inverse design

### 4.1 Large NA lens

Lenses with a large NA have a fast-varying phase profile towards their boundaries, which makes the LPA method less accurate. Here we use SCMT to improve the focal efficiency of a radial meta-lens with NA = 0.8 at wavelength = $650nm$. The metasurface is built with TiO2 square waveguides on a SiO2 substrate, with a period of 400nm. The heights of the individual waveguides are $800nm$, and their widths range from $160nm$ to $380nm$. As a control group, we first designed a meta-lens using the LPA method. Then we improved the focal efficiency of the lens using the SCMT method by initiating with the LPA design in the optimization routine. The object function is to maximize the intensity of the focal spot. The full-wave simulation results of the focal points of the two lenses are shown in figure 4 (a) and (b). The focal efficiency, as defined by the portion of incident power focused inside a circle centering the focal point with a size 3 times the FWHM of the focal point, is improved from 35% to 40%, while the Strehl ratio, obtained by taking the ratio of the peak intensity of the design lens and the ideal lens, is improved from 58% to 66%.

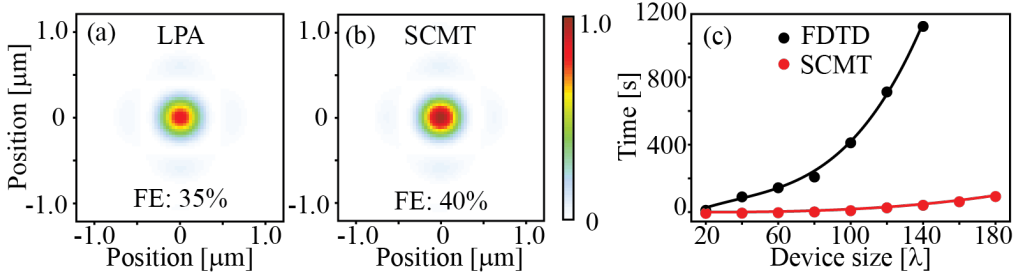

Figure 4. The focal efficiency comparison of two meta lens validated by FDTD simulation. (a) The focal spot of metalens designed by LPA, (b) The focal spot of meta lens designed by SCMT. The full-width-half-maximum (FWHM) for two lenses equal to $0.72\lambda$. The Strehl ratio for the LPA lens is 0.58, and the Strehl ratio for the SCMT lens is 0.66. (c) Comparison of the computing efficiency between SCMT and the FDTD solver (Tidy3d). Both methods are used to simulate 3D metalens with different sizes.

### 4.2 Chromatic aberration reduced lens

Similar to high NA meta-lenses, meta-lenses focusing light at multiple wavelengths also tend to make the LPA method less accurate. To show the improvements SCMT provides over LPA, we use two methods to design meta-lenses that provide focusing for two wavelengths at $650nm$ and $450nm$ simultaneously. With a NA of 0.4, these metasurfaces are also made with TiO2 square waveguides on SiO2 substrate, with a period of 280nm. The height of the waveguide is $800nm$, with widths ranging from $80nm$ to $240nm$. The FDTD validation of two designs at two different wavelengths is shown in figure 5. Although the SCMT-designed lens is a little less efficient than LPA at $650nm$ (figure 5 (a) and (c)), it is much more efficient than LPA at $450nm$ (figure 5 (b) and (d)). The focal efficiency for SCMT designed lens is 42% at $650nm$ and 32% $at$ $450nm$. The focal efficiency for LPA designed lens is 47% at $650nm$ and 22% $at$ $450nm$.

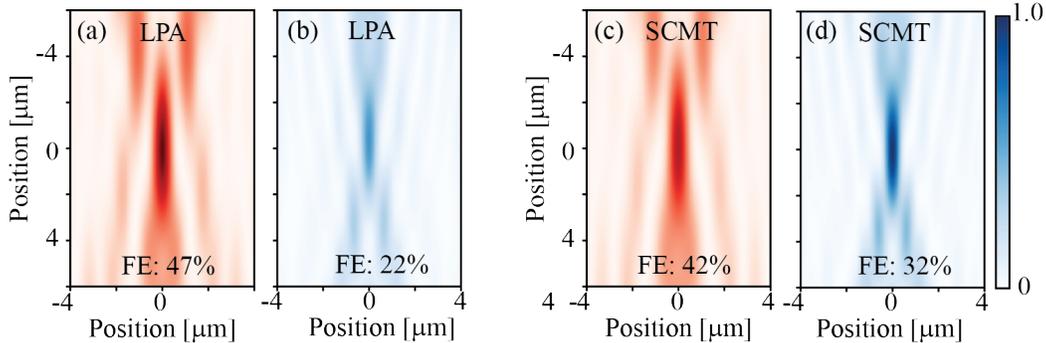

Figure 5. The focal efficiency comparison of LPA and SCMT on design multi-wavelength lens. We use two methods to design lenses that support two wavelengths $650nm$ and $450nm$. The far field of two designed lenses are simulated using the FDTD solver. (a) (b) the focal spot of LPA designed lens at $650nm$ and $450nm$. (c) (d) the focal spot of SCMT designed lens at $650nm$ and $450nm$. (FE: focal efficiency).

### 4.3 Coma aberration reduced lens

Conventionally, coma aberration is suppressed with multi-element optical systems. Metasurface doublet has also been proposed to suppress coma aberration [17], [18], which increases the cost and induce alignment issue. So, a single-layer meta lens with less coma aberration is preferred. However, this situation requires the metasurface to be modeled accurately for plane waves with different incident angles, which cannot be achieved with the LPA method. By equally maximizing the focal intensity for different incident angles using the SCMT model, we design a singlet metalens with greatly suppressed coma aberration at incidence angles up to +/-40 degrees as shown in figure 6. Figure 6 (a) and (c) shows the focal spots of an ideal flat lens that can perfectly focus a normal incident plane wave. However, we can see that as the incident angle increases so does coma aberration. In comparison, the performance of SCMT-designed meta lens is shown in (b) and (d). To our knowledge, this singlet version has yet to be envisioned by the meta-surface community.

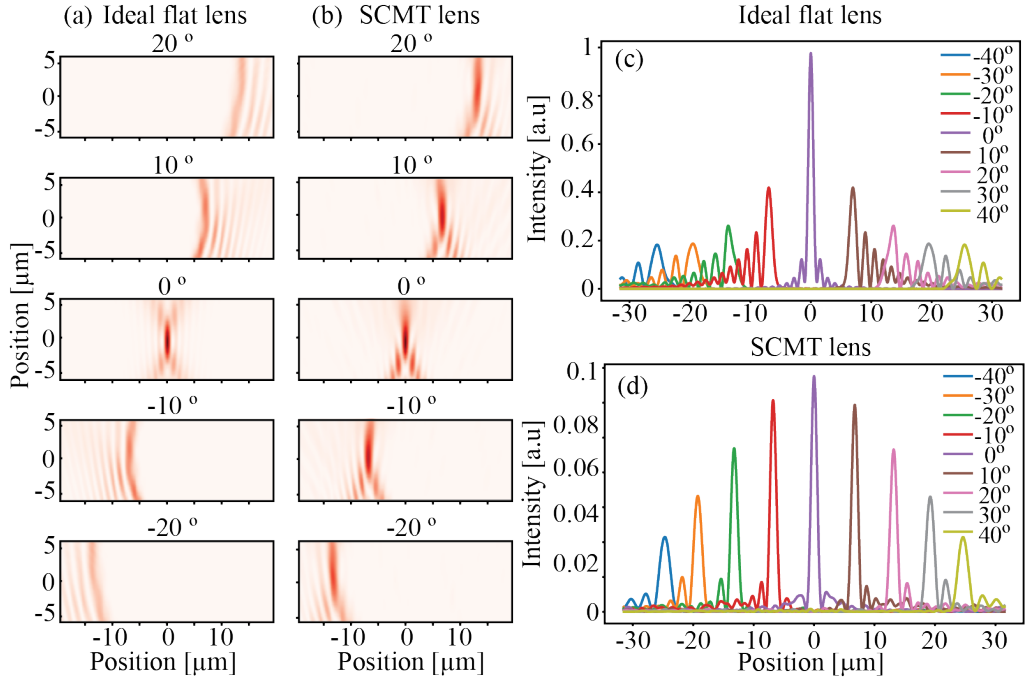

Figure 6. (a) the side view of the focal spot for an ideal flat lens with different incident angles. (b) the side view of the focal spot for SCMT optimized lens with different incident angles. (c) the intensity at the focal plane for an ideal flat lens with different incident angles. (d) the intensity at the focal plane for SCMT optimized lens with different incident angles. The plots are simulated by FDTD.

## 5. Computational efficiency

Finally, we compare our solver's efficiency with the state of art FDTD solver (Tidy3d [19]). Both methods are used to simulate 3D metalens with different sizes, and their time complexities are evaluated. As shown in figure 4 (c). Tidy3d FDTD solver is one of the fastest of its kind, capable of simulating the focusing effect of a $100\lambda \times 100\lambda$ 3D metasurface within 10min. Our solver is at least 10 times faster than tidy3d owing to both highly efficient simulation algorithm and multi-GPU acceleration. To simulate the $180\,\lambda$ x $180\,\lambda$ metasurface, we use 4 NVIDIA TITAN Xp GPUs each with 12GB RAM.

## 6. Discussion

SCMT has been widely used to study and design multi-waveguide systems. The simplest and most practical case is studying the directional coupler, a two-waveguide system, which can be used as power splitters, optical switches, and wavelength filters [20]–[23]. Besides, SCMT has also been used in systems with multiple waveguides, such as multi-core fibers [24], laser arrays [25], and micro combs [26]. Compared with full-wave numerical simulations, SCMT provides a significant speedup in the device design and optimization Because the SCMT-based approach only needs the electromagnetic fields from individual components [27]–[29]. Despite the wide application of SCMT, to our knowledge, it has not been used to design metasurfaces which typically consist of tens of thousands of coupled waveguides. Our study shows that the SCMT is also accurate for modeling multi-waveguide systems at a large scale.

The primary complexity of inverse design problems lies in the calculation of the derivatives to the hyperparameters given an object function. There are three mainstream methods used by the inverse design community. The straightforward approach is the finite-difference method [30], which perturbs each input by a small amount and measures the resulting output. This method is suitable for problems with large number of output but small number of parameters. For forward problems modeled by the Maxwell equation, the adjoint method [31] is usually used, which gets the derivatives by running the forward simulation again with adjoint sources. In the contrary, this method is more efficient for problem with large number of parameters but small number of output. What we used here is automatic differentiation (AD) [32]. AD is the dominant method to calculate derivatives in the machine learning community because it is very efficient to calculate the gradient of a scalar-valued objective with a large number of parameters. In this approach, only the forward simulation is manually implemented while the backward gradient will be generated automatically. However, the constraint is the forward implementation should be differentiable known as differentiable programming. the whole computation can be represented as a finite set of elementary operations of which derivatives are known. In our model, given pre-calculated modal properties, SMCT-based forward simulation is differentiable. However, the process to obtain modal properties through geometrical parameters is not differentiable, leaving a missing link of a fully autonomous inverse design paradigm. To overcome this problem, we replace the eigenmode-solving procedure with a pre-trained neural network as shown in figure 3. This strategy provides a way to convert a non-differentiable model to a differentiable model while maintaining substantially high accuracy. The same idea has been used in the Molecular dynamics simulation community [33], where the neural network is used to fit the potential energy surface of atoms.

In summary, inspired by [8] using temporal CMT to model the metasurface based on resonance, we apply the SCMT to improve the accuracy and the efficiency of the metasurface simulation based on guidance. Our work extends the application of the coupled mode theory in metasurface modeling. Furthermore, we show the accuracy of our method by comparing it with the widely used LPA method and full wave simulation. Finally, an inverse design paradigm based on auto-differentiation is proposed and validated, and an increase of performance is observed. Our method represents a simplified design process capable of realizing complex functionalities.

There are several aspects where this method can be improved. Firstly, our method is not good enough for amplitude response prediction, as shown in figure A.3. Relatively low transmission efficiency compared with conventional optical elements with anti-reflection coatings is one of the key drawbacks constraining metasurfaces from achieving complex functionalities, which typically require propagation through multiple optical elements. Accurately predicting the amplitude response with high efficiency is required to optimize the transmission efficiency of the metasurface. Secondly, our framework requires the designed meta-units to lie on a periodic grid, which limits the fully ability of metasurface. Several works [34], [35] have proved that this limitation can be overcome by topology optimization. Further work needs to be done to incorporate the SCMT method into the topology optimization framework.

## 7. Data availability

The programs that are used to generate all the results are publicly available at https://github.com/airPeter/Meta_SCMT.

## 8. Acknowledgments

The authors acknowledge the support from National Science Foundation (grant no. QII-TAQS-1936359 and no. ECCS-2004685).

## 9. Appendix

### 9.1 Coupled modes of waveguides array

In this section, we will describe the coupled mode propagation in an array of arbitrary waveguides. For an isolated waveguide, a guided mode propagates along its longitudinal direction and retains a constant transverse electromagnetic mode profile. The propagation constant $\beta$ and mode profile can be obtained by solving the eigenmodes of the waveguide. In principle, the same mode analysis can be done for the entire array to obtain the super modes (i.e., eigenmodes of the entire waveguide array) and their propagation constants. To simplify the calculation, the super modes are approximated as linear combinations of the isolated waveguide modes. The role of SCMT is to derive the coefficients of such combinations and their propagation constants. The thorough derivation can be found in [10]. We summarize the results here.

To simplify the illustration, we assume that for an isolated waveguide $i$, the dielectric waveguide supports a single-guided mode with transverse field components $\{E_t^i(x, y), H_t^i(x, y)\}$ and propagation constant $\beta$. (In practice, the meta unit size is always significantly smaller than the wavelength and usually no more than 2 modes will be supported.) Assuming there are $N$ waveguides in the array, for an arbitrary field propagating in the array, it can be approximately decomposed into the isolated waveguide modes based on our assumption. The field profile can be expressed as:

$$E(x,y,z)_t = \sum_{i=1}^{N} u_i(z) E_t^i(x,y) \tag{A.1}$$

$$H(x,y,z)_t = \sum_{i=1}^{N} u_i(z) H_t^i(x,y) \tag{A.2}$$

Where the mode amplitude at position $z$ for mode $i$ is $u_i(z)$. The dynamics of the coefficient vector $U(z) = [u_1(z), u_2(z), ... u_N(z)]$ can be described by the equation:

$$C \frac{dU(z)}{dz} = i(BC + K)U(z) \tag{A.3}$$

Where the $C, K$ are coupling matrices, and $B$ is a diagonal matrix of propagation constants ($B_{ii} = \beta_i$). The element $c_{pm}$ of matrix C is the modal overlapping of the E field of mode $p$ and the H field of the mode $m$:

$$c_{pm} = 2\hat{z} \cdot \iint_{-\infty}^{\infty} E_t^p \times H_t^m dxdy \tag{A.4}$$

And the mode amplitude is normalized such that $C_{pp} = 1, \forall p$.

The element $k_{pm}$ of matrix K is the coupling coefficients of the E field of mode $p$ and the E field of the mode $m$ considering the perturbation to dielectric constant:

$$k_{pm} = \omega \iint_{-\infty}^{\infty} \Delta \epsilon^p [E_t^p \cdot E_t^m] dxdy \tag{A.5}$$

Where $\Delta \epsilon(x,y) = \epsilon(x,y) - \epsilon^p(x,y)$, the $\epsilon(x,y)$ is the dielectric constant of the entire waveguide system. $\epsilon^p$ is the dielectric consisting of isolated waveguides p.

Noticed that the matrix $B, C, K$ are constant once the waveguide array is fixed. Their values can be pre-calculated for a meta-unit library and parameterized based on the meta-unit geometry. By solving equation (1), we can get the coefficient vector $U(z)$ given an input $U(0)$:

$$U(z) = exp\, exp\, [iC^{-1}(BC+K)z]\, U(0) \tag{A.6}$$

Equation (6) is what we need to build our forward solver. Although we don't need the super mode propagation constant $\sigma$ and the super mode coefficient vector $V$, for completeness, the eigenvalue equation used to calculate them is:

$$[C^{-1}(BC+K)]V = \sigma V \tag{A.7}$$

### 9.2 *Free space and waveguides coupling*

To calculate $U(0)$, a straightforward way is to decompose the incident field to the bases formed by the super modes. However, to avoid calculating the super mode, $u_i(0)$ is approximated as the coupling coefficient between the incident field and the isolated waveguide $i$ [36]:

$$u_i(0) = \epsilon_0 c \frac{2 n_0 n_{eff}^i}{n_0 + n_{eff}^i} \iint E_{incident} E_t^i(x,y) dxdy \tag{A.8}$$

Where the effective refractive index $n_{eff} = \frac{\beta}{k}$, and $k = \frac{2\pi}{\lambda}$. $\lambda$ is the wavelength, $\epsilon_0$ is the vacuum permittivity and $c$ is the speed of light in the vacuum.

The transmission coefficient for a single waveguide $i$ is:

$$\eta_i = \frac{n_0 n_{eff}^i}{n_0 + n_{eff}^i} \tag{A.9}$$

The output field is approximated as:

$$E(x,y,h)_t = \sum_{i=1}^{N} \eta_i\, u_i(h) E_t^i(x,y) \tag{A.10}$$

Where $u_i(h)$ is calculated by equation (6) with the notation $h$ as the length of the waveguide.

### 9.3 *free space propagation*

We use the Rayleigh-Sommerfeld diffraction theory [12] to model the light propagating in free space.

$$E_{far}(x,y) = \frac{z}{r^2}\left(\frac{1}{2\pi r} + \frac{1}{j\lambda}\right) exp\left(\frac{j2\pi r}{\lambda}\right) E_{near}(x',y') \quad (A.11)$$

where r is the distance between points $(x', y')$ and $(x, y)$, $z$ is the propagation distance.

### 9.4 *Accuracy comparison between LPA, SCMT, and no coupling*

we compare the accuracy of LPA, SCMT, and no coupling at different coupling strengths. The definition of coupling strength is provided in Appendix 9.5. Basically, the closer the meta units are the stronger their coupling strength is. In figure A.1 (a), 3 methods are used to predict the near-field phase of the metalens under different coupling strengths. The relative error of SCMT and no coupling method are plotted such that the error of the LPA method to the FDTD simulation is 100%. We can see that the SCMT is more accurate than LPA at a strong coupling regime. And 3 methods have similar accuracy at weak coupling regimes. The relative error of SCMT and no coupling amplitude prediction are plotted in figure A.1 (b). We can also see the pattern that SCMT is more accurate at a strong coupling regime. (The LPA is usually not used to predict amplitude, since its prediction is almost constant across the whole metasurface). The definition of the relative error is provided in Appendix 9.6.

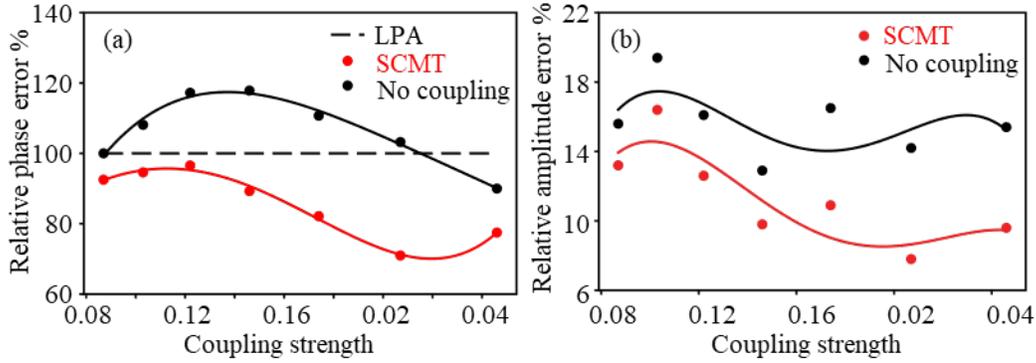

Figure A.1. The accuracy analysis of the SCMT method was validated by FDTD. (a), 3 methods are used to predict the near-field phase of metalens under different coupling strengths. The relative error of SCMT and no coupling method are plotted such that the error of the LPA method to the full-wave simulation is 100%. (b) The relative error of SCMT and no coupling amplitude prediction under different coupling strengths.

### 9.5 *Definition of coupling strength*

For 2 waveguide coupling, the coupling strength can be defined by the overlapping of the mode field:

$$c_{12} = 2\,\hat{z} \cdot \iint_{-\infty}^{\infty} E_t^1 \times H_t^2 \, dxdy \quad (A.12)$$

$$coupling\ trength = \frac{c_{12} + c_{21}}{2} \quad (A.13)$$

For the waveguide array, we can define the coupling strength in the same way. Assuming the meta units are waveguides with square cross sections. For a given period, the allowed waveguide width range is $[w_{min}, w_{max}]$. After choosing a proper step size $dw$ based on the requirements of resolution, The number of possible widths $N_c = \frac{w_{max} - w_{min}}{dw}$. So, the number of different nearest-neighbor coupling combinations is $N_c^2$. The coupling strength for the above-defined system is:

$$coupling\ trength = \frac{\sum_{i=1}^{N_c} \sum_{j=1}^{N_c} c(w_i, w_j)}{N_c^2} \quad (A.14)$$

### 9.6 *Definition of the relative error*

The absolute phase error is defined as:

$$ape(p1, p2) = \left\|e^{ip1} - e^{ip2}\right\|_1 \quad (A.15)$$

Here, we use $L_1$ norm to increase the robustness of the error to outliers.

The relative error of the SCMT method is defined as:

$$rpe_{SCMT} = \frac{ape(p_{SCMT}, p_{FDTD})}{ape(p_{LPA}, p_{FDTD})} \quad (A.16)$$

$p_{SCMT}$ is phase prediction given by the SCMT method. We can define the relative error of the no-coupling method in the same way.

For the relative amplitude error, the definition is straightforward:

$$rae_{method} = \frac{||amp_{method} - amp_{FDTD}||^2}{||amp_{FDTD}||^2} \quad (A.17)$$

### 9.7 Neural network submodule

For the submodule that output $c_{ij}$, the input of the network is: the distance between waveguide $i$ and waveguide $j$ (for the 2D model, the distance is encoded by 1 number $x$, for the 3D model, the distance is encoded by a tuple $(x, y)$), the mode index $m_i$ and $m_j$ (e.g. for the fundamental mode of the0 waveguide $m = 0$), and the waveguide widths $h_i, h_j$.

The number of the training dataset equals the total combination of the input. Taking a 2D system as an example. The number of possible widths is $N_h$, the number of possible modes is 2, and the number of possible distances is 5 ({-2, -1, 0, 1, 2} times period). So, the size of the training dataset is $10N_h^2$. We train the fully connected network so that the relative error of the model prediction and the ground truth is smaller than 0.1%.

For the submodule that output $k_{ij}$, from equation (5), we notice that $k_{ij}$ is not only dependent on waveguides $i$ and $j$, but also the waveguides in between (coming from the term $\Delta\epsilon$). In practice, the error of ignoring the waveguides in between is < 1%. So we define the submodule input the same as the $c_{ij}$ submodule.

For the submodule that output $E_i$, the input is the mode index $m_i$ and the waveguide width $h_i$. The output is a discretized field distribution (a matrix for the 3D model and an array for the 2D model).

### 9.8 Calculate $C^{-1}$ with approximation

Assume a matrix $C$ has size $N \times N$ which uses $O(N^2)$ memory, a standard algorithm to calculate the $C^{-1}$ is doing the LU decomposition of matrix $C$ and then using the Gaussian elimination. The $C^{-1}$ also has size $N \times N$. In our case, the matrix $C$ is a stripped diagonal matrix as illustrated in figure A.2 which can be sparsely represented using $O(N)$ memory. However, if we use a standard algorithm to calculate $C^{-1}$ which will inevitably use $O(N^2)$ memory. To overcome this difficulty, we approximate the $C^{-1}$ row-wise. The process is illustrated in figure A.2. We want to minimize the L2 distance of the identity matrix and the $C^{-1}_{appox}C$ under the constraint that only a band of diagonal elements can be non-zero. (There is a trade-off of the width of the non-zero band between the memory and the accuracy. In our case, we choose the width = $11\sqrt{N}$). This can be converted to a row-wise optimization problem as shown in figure A.2 (b), where $C_s$ is a submatrix of $C$, $C_{is}^{-1}$ is a row of $C_{approx}^{-1}$. $I_s$ is a sub-row of $I$ where only 1 element is non-zero. There is an analytical solution that minimizes the row-wise loss $L_i$:

$$(C_{si}^{-1})^T = argmin(L_i) = (C_s C_s^T)^{-1} C_s I_s^T \quad (A.18)$$

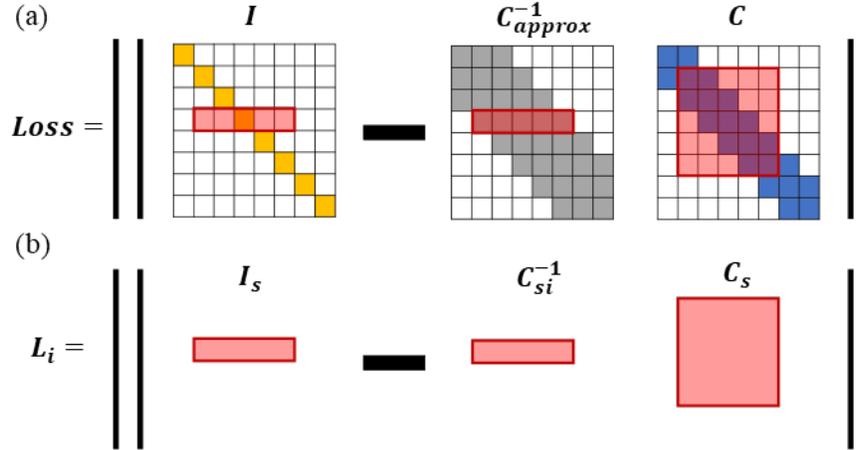

Figure A.2. Diagram of approximate $C^{-1}$. (a) The ideal $C^{-1}_{approx}$ is a matrix that minimizes the L2 loss $\|I - C^{-1}_{approx}C\|$ under the constraint that only a band of diagonal elements can be non-zero. (b) The optimization can be solved row-wisely. $C_s$ is a submatrix of $C$, $C_{is}^{-1}$ is a row of $C^{-1}_{approx}$. $I_s$ is a sub-row of $I$ where only 1 element is non-zero.

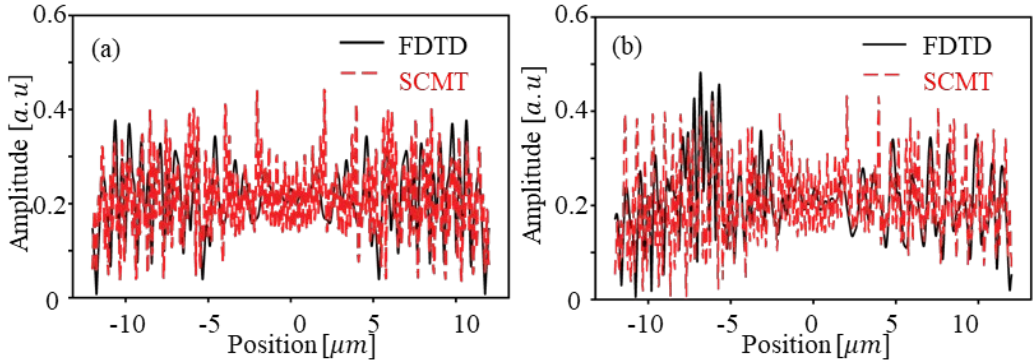

Figure A.3. The accuracy analysis of the SCMT method was validated by FDTD. (a) the near field amplitude comparison of SCMT prediction and the FDTD simulation with a normal incident plane wave. (b) the near field amplitude comparison of SCMT prediction and the FDTD simulation with a 15° incident plane wave.